\documentclass[twocolumn]{jpsj3_miyazaki}
\usepackage{graphicx}
\usepackage{lipsum}
\usepackage{amsmath}
\usepackage{bm}
\usepackage{color}
\usepackage{braket}
\usepackage[dvipsnames]{xcolor}
\usepackage{ulem}
\usepackage{dcolumn}

\def\@jpsjmark{}
\def\@JPSJmark{}

\title{Linear Flavor-Wave Analysis of\\SU(4)-Symmetric Tetramer Model with Population Imbalance}

\author{Yuki Miyazaki$^1$\thanks{y-miyazaki@phys.aoyama.ac.jp},  Giacomo Marmorini$^{1,2}$, Nobuo Furukawa$^1$, and Daisuke Yamamoto$^2$}
\inst{$^1$Department of Physics and Mathematics, Aoyama Gakuin University, Sagamihara, Kanagawa 252-5258, Japan\\
$^2$Department of Physics, College of Humanities and Sciences, Nihon University, Sakurajosui, Setagaya, Tokyo 156-8550, Japan}

\abst{
We study the quantum magnetism of the SU(4) Mott insulator in a square optical superlattice, in which atoms with four nuclear-spin components strongly interact with each other, in the presence of an external field that controls the imbalance between the population of two components and that of the other two. This is a natural extension of the physics of spin-dimer materials under strong magnetic field.
We apply an extended linear flavor-wave theory based on four-site plaquettes and unveil the ground-state phase diagram and excitation spectra. When the population of the four components is balanced and the plaquesttes are weakly coupled, the ground state is approximately given by the direct product of local SU(4)-singlet states. In high-field, the system reaches a ``saturated state'' where only two components are present.
Our main finding is a nontrivial intermediate phase, which has a checkerboard-like arrangement of the SU(4)-singlet and four-site resonating-valence-bond states.}

\begin{document}
\maketitle

Quantum dimer magnets, which consist of a weakly coupled network of $S=1/2$ spin pairs, have been attracting attention as a prototypical example for studying novel phases of matter and quantum phase transitions. Spin-dimer materials such as $\rm{TlCuCl_3}$ \cite{TlCuCl3_Nikuni,TlCuCl3_Yamada,TlCuCl3_Matsumoto}  feature a non-magnetic gapped ground state, which is continuously connected to a direct product of spin-singlet states.
The lowest excitations to one of the spin-triplet states behave like bosonic quasiparticles, called triplons. When a magnetic field is applied, the energy gap decreases and eventually closes at a certain critical field. The resulting quantum phase transition from non-magnetic to magnetic state can be explained by Bose-Einstein condensation of triplons. Some exceptional spin-dimer materials with strong frustration effects, {\it e.g.}, $\rm{Ba_2CoSi_2O_6Cl_2}$ \cite{Ba2CoSi2O6Cl2_Kurita}, which exhibits a crystallized state of localized triplons, and $\rm{SrCu_2(BO_3)_2}$ \cite{SrCu2BO32_Koga}, characterized by orthogonal dimers forming a Shastry-Sutherland lattice, have also been studied with great interest.

In the field of cold atomic and optical physics, a great deal of effort is being made to explore exotic quantum magnetism using ultracold atomic gases loaded in an optical lattice \cite{Coldatom_Pagano,Coldatom_Scazza,Coldatom_Zhang}. One of the recent remarkable achievements is the observation of long-range  antiferromagnetic correlations of an artificial antiferromagnet realized with the two lowest hyperfine states of the fermionic species ${}^6\rm{Li}$ \cite{SU2_Mazurenko}. 
Besides the conventional SU(2) systems, there has been success in realizing higher symmetric Mott insulators with alkaline-earth(-like) atoms such as ${}^{173}\rm{Yb}$ \cite{SUN_Mott_Taie,SUN_Mott_Hofrichter}. Those atoms are closed-shell in the ground state, and the nuclear spin provides $2I+1$ states (flavors), where $I=5/2$ for ${}^{173}\rm{Yb}$. With these atomic species one can prepare ideal SU($\mathcal{N}>2$)-symmetric systems because the interatomic interaction does not depend on the nuclear spin to a great approximation \cite{SUN_scatteringlength}. Quantum magnetism with a high SU($\mathcal{N}$) symmetry has great potential to feature exotic phase transition phenomena associated with a rich variety of spontaneous symmetry breaking.

\begin{figure}[h]
    \centering
    \includegraphics[width=5cm]{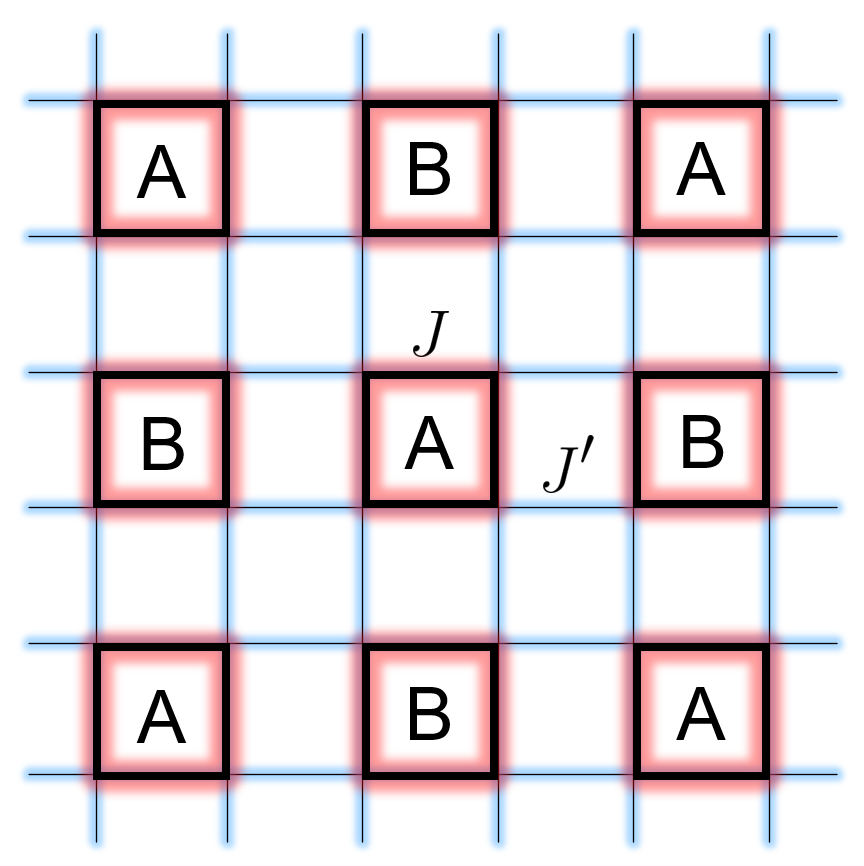}
    \caption{Sketch of the tetramerized square lattice with the assumed two-sublattice structure. The line thickness represents the strength of the interactions; $J\gg J'>0$. }
     \label{f1}
\end{figure}

Here, we consider the square superlattice in which the lattice sites are tetramerized as shown in Fig. 1 and focus on the case where each of the sites is singly occupied by one of four pseudospin components (flavors) of fermions with the SU(4) symmetry.  To study the strong coupling physics, we describe the system by an extended antiferromagnetic Heisenberg model with each site comprising the fundamental representation of the SU(4) group.

In cold-atom experiments, such an optical superlattice can be created by superimposing long-period and short-period optical square lattices \cite{superlattice}. Progress in loading the above-mentioned ytterbium isotope in this kind of superlattice has already been reported. \cite{takahashi-20} In the field of conventional  SU(2) magnetism, analogous plaquette-based models have been  considered\cite{tetra_Zhitomirsky,tetra_Ueda}. Let us also mention that  SU(4) symmetry has seen a recent revival in models of spin-orbital physics in solid state compounds, {\it e.g.} $\alpha$-ZrCl$_3$.\cite{yamada-18}

\begin{figure}[h]
    \centering
    \includegraphics[width=9cm]{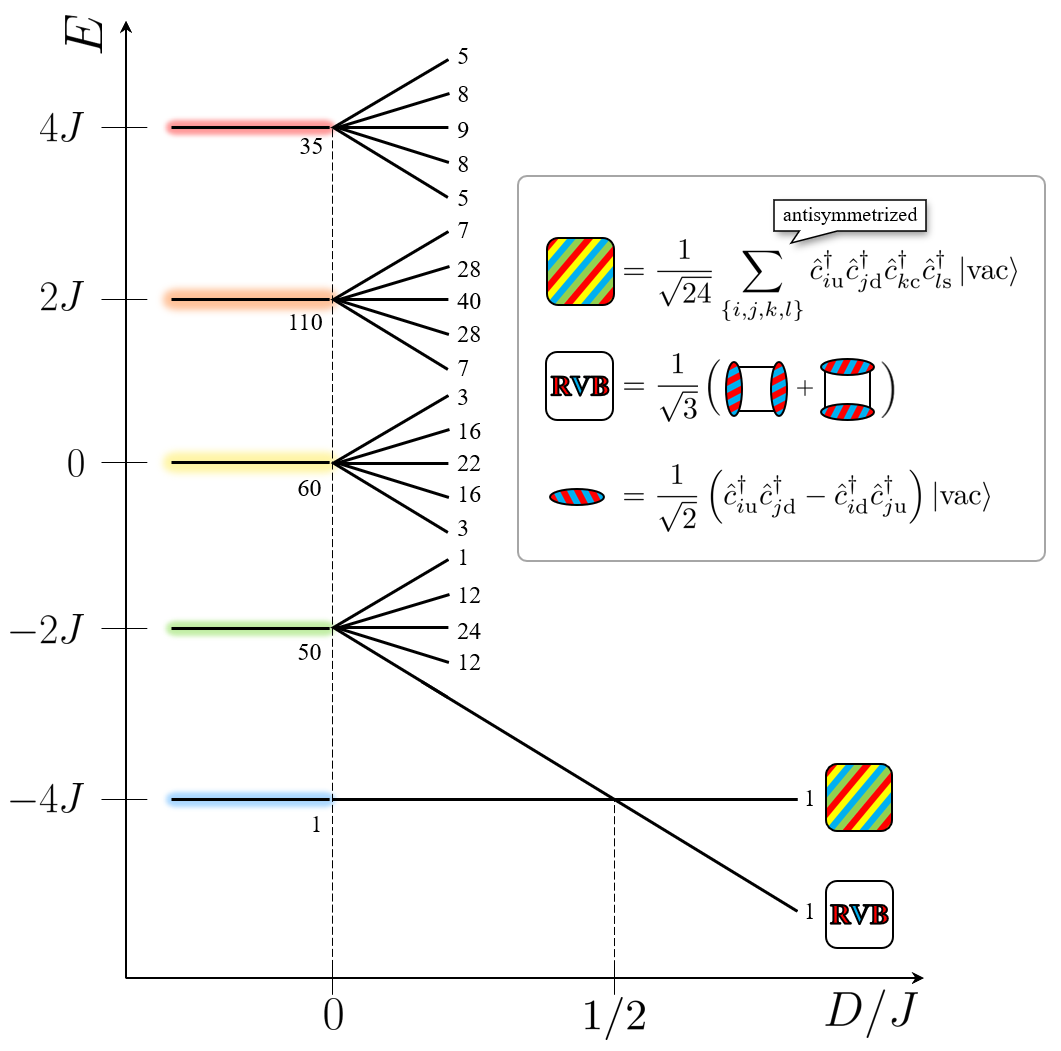}
    \caption{The energy levels of an isolated plaquette described by the SU(4) Heiseneberg {model under an external field that controls the population imbalance between two components and the other two}. The numbers {attached} to each level are the degree of degeneracy.}
   \label{f2}
\end{figure}

Hereafter, let us name the four flavors ``u", ``d", ``c", and ``s" after the quark model.
The SU(4) Heisenberg model with a multiplet in the fundamental representation at each site is given by the Hamiltonian \cite{LFWT_SU3_1,SUN_swapping,SU6_swapping,dimerNeel}
\begin{eqnarray}
\hat{\mathcal{H}}_{\rm{SU(4)}}&=&\sum_{{\mathcal{h}i,j\mathcal{i}}}\sum_{\mu,\nu}J_{i,j}\hat{F}^{\nu}_{\mu}(i)\hat{F}^{\mu}_{\nu}(j)
\end{eqnarray}
where $J_{i,j}$ are superexchange couplings; they take the value $J(>0)$ within a plaquette and $J'(>0)$ between neighboring plaquettes. We define the flavor-changing operator $\hat{F}^{\nu}_{\mu}(i)\equiv\mathcal{j}\mu(i)\mathcal{i}\mathcal{h}\nu(i)\mathcal{j}$, where $\mu, \nu\in \{$u, d, c, s$\}$. For an isolated plaquette ($J'=0$), there are five energy levels: the lowest one is non-degenerate and corresponds to the SU(4) singlet state, namely the fully antisymmetrized state $\frac{1}{\sqrt{24}}\sum_{\left \{ i, j, k, l \right \}}\hat{c}_{i\rm{u}}^{\dagger}\hat{c}_{j\rm{d}}^{\dagger}\hat{c}_{k\rm{c}}^{\dagger}\hat{c}_{l\rm{s}}^{\dagger}\mathcal{j}\rm{vac}\mathcal{i}$, where $\sum_{\left \{ i, j, k, l \right \}}$ denotes the summation over the 24 possible permutations of site indexes $i,j,k,$ and $l$
and $\hat{c}_{i\mu}^\dagger$ is the creation operator of a fermion at site $i$ with flavor $\mu$ \cite{SU4singlet}. 
Each of the other four levels is massively degenerate. In order to consider a natural extension of the spin-dimer physics to SU(4), we apply an external field which controls the population imbalance between $\{$u, d$\}$ and $\{$c, s$\}$ and study the Hamiltonian  $\hat{\mathcal{H}}=\hat{\mathcal{H}}_{\rm{SU(4)}}+\hat{\mathcal{H}}_D$, where
\begin{equation}
    \hat{\mathcal{H}}_D=D\sum_i\left[\hat{F}^{\rm{c}}_{\rm{c}}(i)+\hat{F}^{\rm{s}}_{\rm{s}}(i)-\hat{F}^{\rm{u}}_{\rm{u}}(i)-\hat{F}^{\rm{d}}_{\rm{d}}(i)\right].
\end{equation}
When the value of $D(>0)$ is increased, the population of $\{$u, d$\}$ tends to increase while $\{$c, s$\}$ components are disfavored. 
This presents as a natural analogy with the conventional magnetic field for SU(2) spins
in terms of dividing the single-particle energy levels into two levels symmetrically.
For the case of isolated tetramers ($J'=0$), the transition from the SU(4)-singlet state to the four-site resonating-valence-bond state of u and d (RVB${}_{\rm{ud}}$) occurs via the energy level crossing at $D/J= 1/2$ as shown in Fig. \ref{f2}. In previous works, the four-site RVB state was created using two atomic states of ${}^{87}$Rb loaded into a {plaquette of an optical lattice}, and observed by the singlet-triplet oscillation technique {\cite{RVB_Belen,RVB_Nascimbene}}.
For finite $J'(>0)$, the appearance of a nontrivial intermediate phase between SU(4) singlet phase and RVB${}_{\rm{ud}}$ phase is expected, in analogy with spin dimer magnets. 

We study the ground state of the Hamiltonian $\hat{\mathcal{H}}$ using an extended linear flavor-wave theory (LFWT) \cite{LFWT_form_1,LFWT_form_2,LFWT_form_3,LFWT_SU3_1,LFWT_SU3_2,LFWT_SU4_SU6} based on four-site plaquettes. First, we reduce the full problem to a four-site one by applying the mean-field approximation $\hat{F}^{\nu}_{\mu}(i)\hat{F}^{\mu}_{\nu}(j)\approx
\mathcal{h}\hat{F}^{\nu}_{\mu}(i)\mathcal{i}\hat{F}^{\mu}_{\nu}(j)+
\mathcal{h}\hat{F}^{\mu}_{\nu}(j)\mathcal{i}\hat{F}^{\nu}_{\mu}(i)-
\mathcal{h}\hat{F}^{\nu}_{\mu}(i)\mathcal{i}\mathcal{h}\hat{F}^{\mu}_{\nu}(j)\mathcal{i}$ to the interplaquette bonds; this approach, named plaquette mean-field (PMF) method, is clearly justified for small $J'$. We calculate the mean-field parameters $\{ \mathcal{h}\hat{F}^{\nu}_{\mu}(i)\mathcal{i}  \}$ self-consistently via the solution of the four-site problem under the assumption of the plaquette-checkerboard pattern shown in Fig. \ref{f1} \cite{CMF_EX1,CMF_EX2,CMF_EX3}. {Subsequently, we include the interplaquette quantum correlation around the PMF ground state}. To this end, we rewrite the Hamiltonian $\hat{\mathcal{H}}$ in terms of the creation (and annihilation) operators of the ground and excited states of the PMF Hamiltonian, $\hat{g}_I^\dagger$ and $\{\hat{e}_{I,n}^\dagger\}$, where $I$ denotes the plaquette index and $n$ labels the 255 excited states. {Those operators obey the bosonic commutation relations under the constraint   
\begin{equation}
 \hat{g}_I^\dagger\hat{g}_I+\sum_{n}\hat{e}_{I,n}^\dagger\hat{e}_{I,n}=1.
 \end{equation}
 }
Following the analogy with the spin-wave  method\cite{Spinwave_Anderson,Spinwave_Manousakis,Spinwave_Zhitomirsky}, we may expand the Hamiltonian with 
 \begin{equation}
 \hat{g}_I^\dagger=\sqrt{1-\sum_{n}\hat{e}_{I,n}^\dagger\hat{e}_{I,n}}\simeq1-\frac{1}{2}\sum_{n}\hat{e}_{I,n}^\dagger\hat{e}_{I,n}+\cdots,
 \end{equation}
 under the assumption that the fluctuations  $\sum_n\hat{e}_{I,n}^\dagger\hat{e}_{I,n}$ around the PMF ground state are sufficiently small; this can be verified {\it a posteriori} in a large region of the parameter space (see the Supplementary Material\cite{SM}).
The Hamiltonian becomes
\begin{equation}
    \hat{\mathcal{H}}=E_0+\hat{\mathcal{H}}_1+\hat{\mathcal{H}}_2+\hat{\mathcal{H}}_3+\cdots,
\end{equation}
where $E_0$ is the energy within the PMF approximation and $\hat{\mathcal{H}}_n$ denotes the $n$-th order correction in $\hat{e}_{I,n}^\dagger,\hat{e}_{I,n}$. Note that $\hat{\mathcal{H}}_1=0$ is satisfied. Performing the Fourier transformation ($\hat{e}_{I,n}\rightarrow \hat{e}_{\mbox{\boldmath$k$},n}$) and the diagonalization via the generalized  Bogoliubov transformation ($\hat{e}_{\mbox{\boldmath$k$},n}\rightarrow \hat{\varepsilon}_{\mbox{\boldmath$k$},\lambda}$) \cite{Bogoliubov_Colpa}, we obtain 
\begin{equation}
    \hat{\mathcal{H}}_2=\sum_{\mbox{\boldmath$k$}}\sum_{\lambda=1}^{510}\omega_{\mbox{\boldmath$k$},\lambda}\hat{\varepsilon}^\dagger_{\mbox{\boldmath$k$},\lambda}\hat{\varepsilon}_{\mbox{\boldmath$k$},\lambda}+E_{\rm{zp}},
\end{equation}
where $\omega_{\mbox{\boldmath$k$},\lambda}$ is the $\lambda$-th excitation energy band and 
\begin{equation}
E_{\rm{zp}}=\frac{1}{2}\sum_{\mbox{\boldmath$k$}}\sum_{\lambda}\omega_{\mbox{\boldmath$k$},\lambda}-\frac{1}{2}\sum_{\mbox{\boldmath$k$}}\textup{Tr}A_{-\mbox{\boldmath$k$}}^*.
\end{equation}
is the zero-point energy. Here, $A_{-\mbox{\boldmath$k$}}^*$ is the bottom-right $510\times510$ block of the Bogoliubov matrix representation of $\hat{\mathcal{H}}_2$ \cite{Bogoliubov_Colpa,Choleskydecomposition}. See the Supplementary Material\cite{SM} for more technical details.

\begin{figure}[h]
    \centering 
    \includegraphics[width=8.8cm]{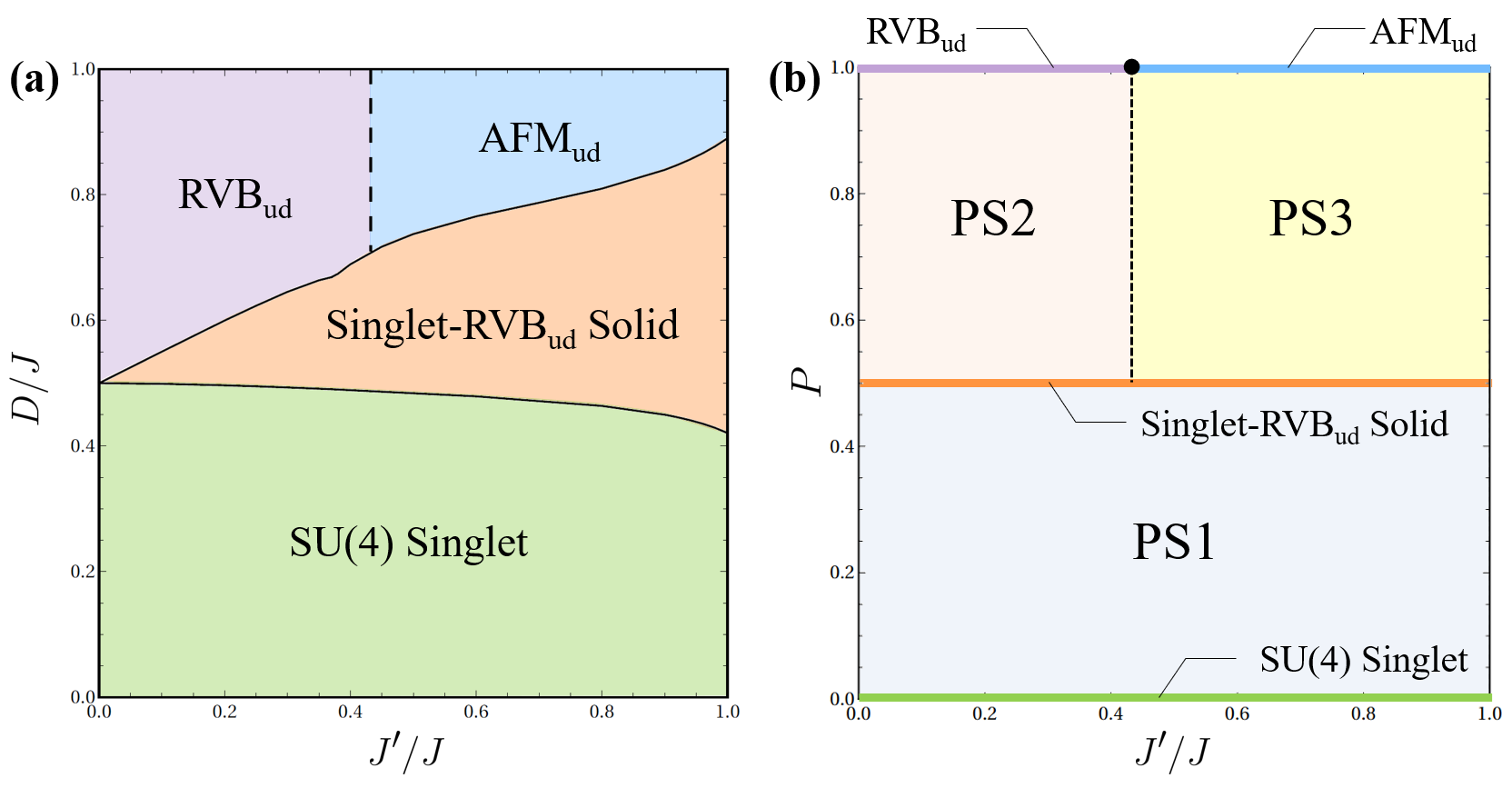}
    \centering
    \includegraphics[width=7.8cm]{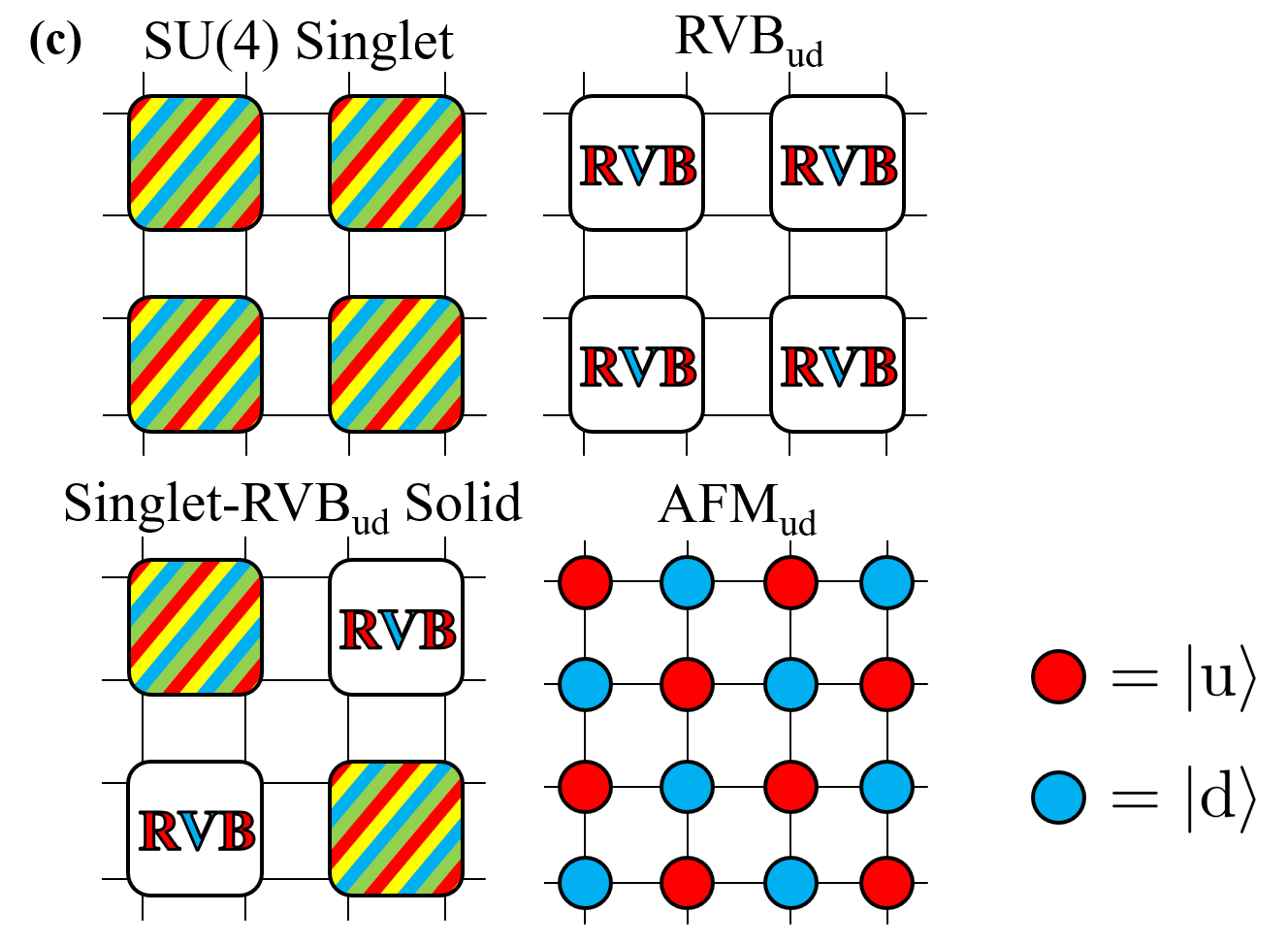}
    \caption{(a) Ground-state phase diagram in the  $J'/J$ vs. $D/J$ plane  within LFWT. Dashed  and solid lines represent  second-order and first-order   boundaries, respectively. (b) The phase diagram in the  $J'/J$ vs. $P$ plane. (c) Sketches of each phase {(see also Fig. \ref{f2}).}}
    \label{f3}
\end{figure}

\begin{figure}[h]
    \centering
    \includegraphics[width=8cm]{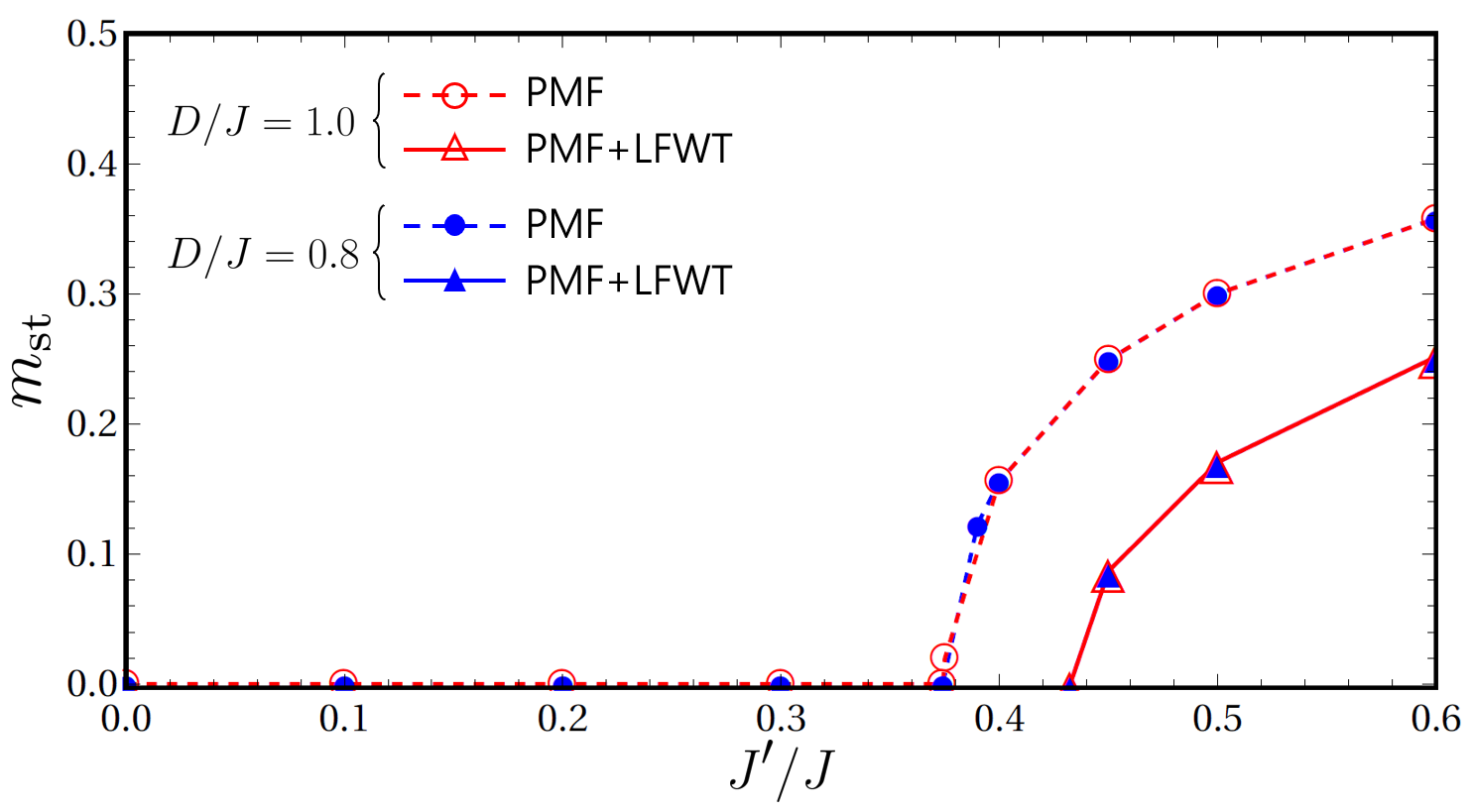}
    \caption{The staggered magnetization for increasing $J'/J$ at $D/J=0.8$ and $D/J=1.0$ {calculated by PMF (circles) and PMF$+$LFWT (triangles).}}
    \label{f4}
\end{figure}

\begin{figure}[h]
    \centering
    \includegraphics[width=6cm]{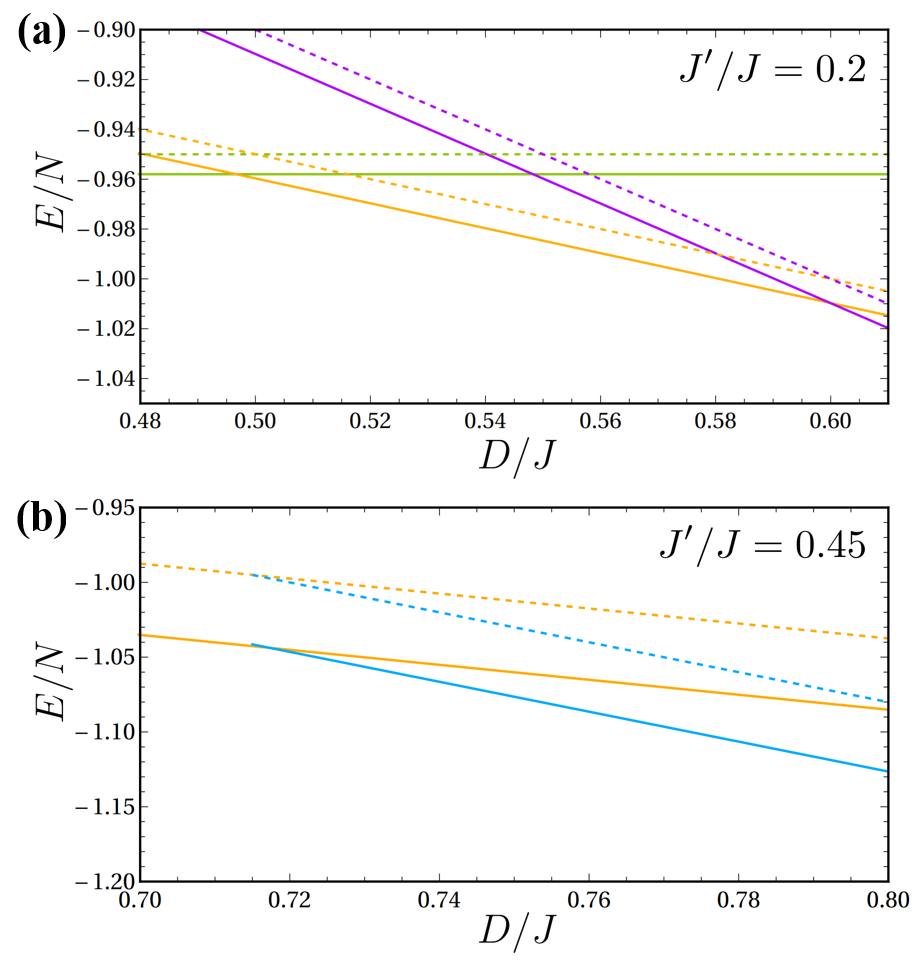}
    \caption{The energy per site as a function of $D/J$ at (a) $J'/J=0.2$ and (b) $J'/J=0.45$. The  color code is as follows: green for SU(4) singlet, orange for SU(4)-RVB${}_{\rm{ud}}$ solid, violet for RVB${}_{\rm{ud}}$, and light blue for AFM${}_{\rm{ud}}$. The dashed  and solid lines represent $E_0/N$ and $(E_0+E_{\rm{zp}})/N$, respectively.}
    \label{f5}
\end{figure}

Figure \ref{f3}(a) shows the ground-state phase diagram in the plane of $J'/J$ vs. $D/J$ within the {PMF$+$}LFWT calculations. We also show the phase diagram in the plane of $J'/J$ vs. $P$ [Fig. \ref{f3}(b)], where $P\equiv-\braket{\partial\hat{\mathcal{H}}/\partial D}/N$ is the conjugate variable of $D$. 
In the regions labeled PS1, PS2, and PS3, phase separation of the two phases which sandwich the respective PS region occurs: SU(4) singlet and singlet-RVB${}_{\rm ud}$ solid phases for PS1, singlet-RVB${}_{\rm  ud}$ solid and RVB${}_{\rm ud}$ phases for PS2, and singlet-RVB${}_{\rm ud}$ solid and AFM${}_{\rm ud}$ phases for PS3. The  phase domains differ by the value of $P$, that can be
rewritten as
\begin{equation}
    P=\frac{\braket{\hat{N}_{\rm{u}}}+\braket{\hat{N}_{\rm{d}}}-\braket{\hat{N}_{\rm{c}}}-\braket{\hat{N}_{\rm{s}}}}{	N},
    \label{opp}
\end{equation}
where $\hat{N}_{\mu}=\sum_i\hat{c}_{i\mu}^\dagger\hat{c}_{i\mu}=\sum_i\hat{F}^\mu_\mu(i)$ is the total number operator of particles {of} flavor $\mu$. The sketches of each phase are depicted in Fig. \ref{f3}(c). Note that our LFWT analysis, being  based on the PMF approximation, can reliably describe the physics at relatively small values of $J'/J$. 

In the SU(4) singlet phase, since the population of the four components is balanced, the value of $P$ is zero. In the RVB${}_{\rm{ud}}$ phase, instead, $P=1$ because both c and s components are absent. In fact, when the value of $D$ is large enough the system can be mapped to a simple SU(2)-symmetric system with only u and d components. When the value of $J'$ is increased in such a regime, a second-order phase transition from RVB${}_{\rm{ud}}$  to N\'eel antiferromagnetic phase of u and d (AFM${}_{\rm{ud}}$) occurs, whose characteristic order parameter is the staggered magnetization. In order to determine the critical point, we consider the Hamiltonian with staggered magnetic field:
\begin{equation}
    \hat{\mathcal{H}}=\hat{\mathcal{H}}_{\rm{SU(4)}}+\hat{\mathcal{H}}_D-h_{\rm{st}}\sum_i(-1)^i\hat{S}_i^z,
\end{equation}
where
\begin{equation}
    \hat{S}_i^z=\frac{1}{2}\hat{F}^{\rm u}_{\rm u}(i)-\frac{1}{2}\hat{F}^{\rm d}_{\rm d}(i)+\frac{3}{2}\hat{F}^{\rm c}_{\rm c}(i)-\frac{3}{2}\hat{F}^{\rm s}_{\rm s}(i).
\end{equation}
The PMF staggered magnetization $m_{\rm{st}}$ and the {LFWT} correction $\delta m_{\rm{st}}$ are given by the thermodynamic relation:
\begin{equation}
   m_{\rm{st}}+\delta m_{\rm{st}}=-\frac{1}{N}\left.\frac{d
   (E_0+E_{\rm{zp}})}
   {d h_{\rm{st}}}\right |_{h_{\rm{st}}=0}.
\end{equation}
The critical line, signaled by the onset of the staggered magnetization, is $(J'/J)_{\rm{c}}\simeq 0.374$ at  the PMF level and $0.432$ within {PMF$+$}LFWT, independently of $D$, as shown in Fig. \ref{f4}. In terms of $P$, the AFM${}_{\rm{ud}}$ phase is also characterized by $P=1$. RVB${}_{\rm{ud}}$ phase and AFM${}_{\rm{ud}}$ phase are divided by the critical point $(J'/J)_{\rm{c}}$ [Fig. \ref{f3}(b)]. 
The most interesting feature of the phase diagram is a nontrivial intermediate phase between SU(4) singlet  and RVB${}_{\rm{ud}}$ phases. It is the checkerboard configuration of SU(4)-singlet and RVB${}_{\rm{ud}}$ plaquettes, which we name ``singlet-RVB${}_{\rm{ud}}$ solid''. Since the number of $\{$u, d$\}$ is twice that of $\{$c, s$\}$, this phase is characterized by $P=1/2$. 
The various first-order phase transitions from this phase to the others are due to the energy level crossings depicted in 
Fig. \ref{f5}.

\begin{figure}[h]
    \centering
    \includegraphics[width=8.5cm]{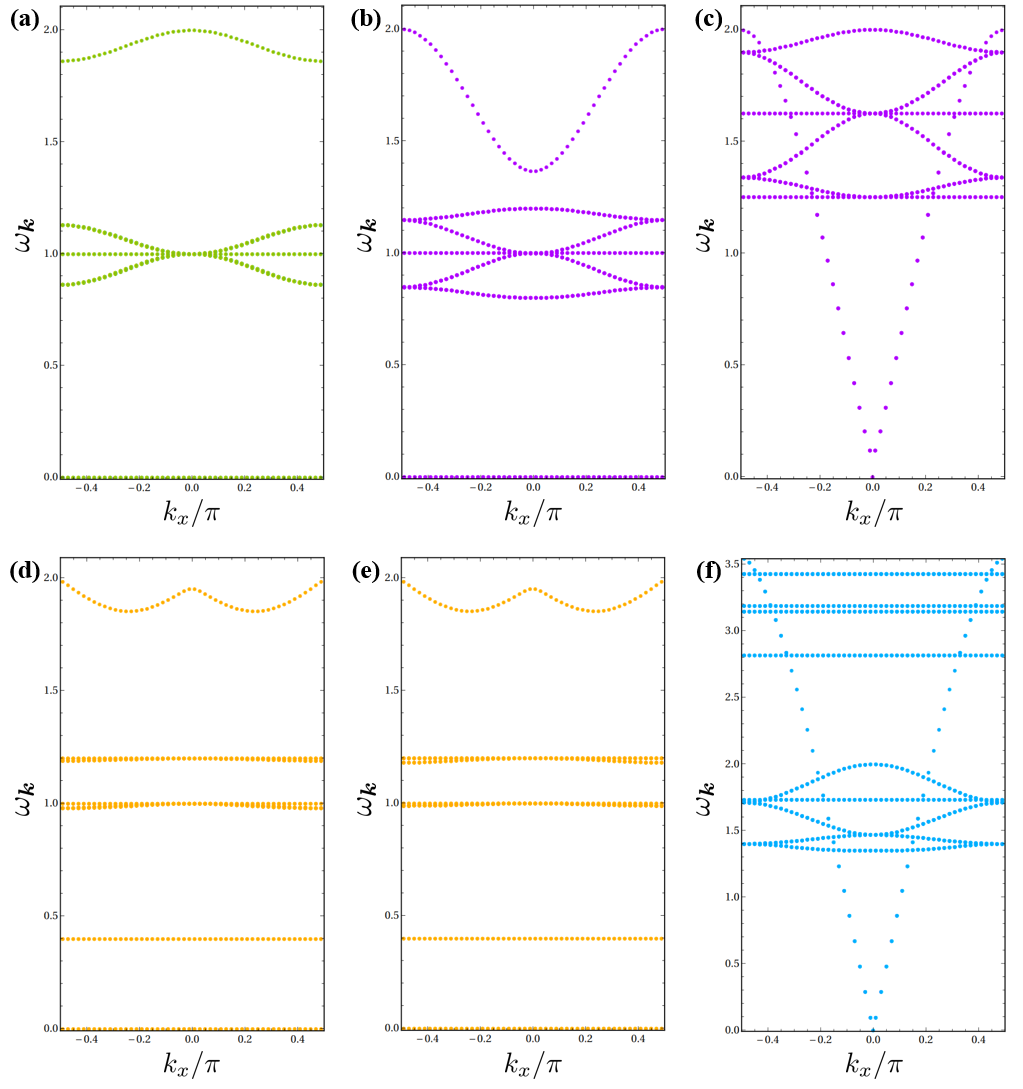}
    \caption{Low-lying excitation spectra calculated from (a) SU(4) singlet phase at $J'/J=0.2, D/J=0.5$, (b) RVB${}_{\rm{ud}}$ phase at $J'/J=0.2, D/J=0.6$ and (c)$J'/J=0.374, D/J=1.0$, (d) Singlet-RVB${}_{\rm{ud}}$ solid phase at $J'/J=0.2, D/J=0.5$ and (e)$D/J=0.6$, and (f) AFM${}_{\rm{ud}}$ phase at $J'/J=1.0, D/J=1.0$. Now the first Brillouin zone is a 
square with vertices $(-\pi/2,0),(0,-\pi/2),(\pi/2,0)$, and $(0,\pi/2)$, but here we show the plots along the $k_x$ axis.}
\label{f6}
\end{figure}

Figures \ref{f6}(a)-(f) show the low-energy excitations for various choices of parameters  within PMF$+$LFWT. At the critical point $(J'/J)_{\rm{c}}\simeq0.374$, there appears a Nambu-Goldstone mode in the LWFT spectrum $\omega_{\mbox{\boldmath$k$}}$,  which is gapless at the $\Gamma$ point [Fig. \ref{f6}(c)]. This is because of the spontaneous symmetry breaking of SU(2) 
down to U(1) in AFM${}_{\rm{ud}}$ phase [Fig. \ref{f6}(f)]. 
 At the PMF phase boundaries between SU(4) singlet and singlet-RVB${}_{\rm{ud}}$ solid phases and between singlet-RVB${}_{\rm{ud}}$ solid and AFM${}_{\rm{ud}}$ phases, there appears a lowest-energy flat band responsible for the gap closing at all wave-vectors $\mbox{\boldmath$k$}$ [Fig. \ref{f6}(a), (b), (d), and (e)]. This arises from the infinite degeneracy of the ground state at the PMF level. However, this will be changed with the inclusion of corrections beyond LFWT, that will give these bands a finite width.
 
 Lastly, let us comment about previous works on the SU(4) Heisenberg model on a uniform square lattice ($D/J=0, J'/J=1$ in our model), where it is suggested that the ground state is the dimer Néel phase \cite{dimerNeel}. In our study, we cannot confirm the stability of the dimer Néel phase at any point in the phase diagram; however, we ascribe this to the limitations of the PMF{$+$LFWT} method, which gives reasonable results at relatively small $J'/J$ but is not suitable for the uniform lattice limit.

 In conclusion, we investigate a natural extension of the spin-dimer materials to SU(4)-symmetric tetramer systems and study the quantum phase transitions in the presence of flavor population imbalance using the {PMF$+$}LFWT.  We  unveil the ground-state phase diagram and show the low-energy excitation spectrum. We find a nontrivial intermediate phase in the saturation process of the order parameter Eq.~\ref{opp}. The Hamiltonian we discussed can be realized by loading ${}^{173}$Yb atoms, whose nuclear spin state populations can be controlled by the optical pumping technique\cite{opticalpumping}, into an optical superlattice\cite{superlattice}. One way to neatly detect the new intermediate phase is to observe atoms on each site directly by quantum-gas microscope\cite{QGM_Bakr,SU2_Mazurenko,QGM_Cheuk,QGM_Haller,QGM_Miranda_1,QGM_Miranda_2,QGM_Yamamoto}; this kind of measurement is, however, still not easy in SU(${\cal N}$) cold atom experiments. On the other hand, probing the system with the singlet-triplet oscillation technique, already successfully used in SU(${\cal N}$) experiments,\cite{opticalpumping} could provide a significant, if indirect, signal.

\begin{acknowledgments}
We thank Y. Takahashi, Y. Takasu, and S. Taie, I. Danshita, and S. Tsuchiya for useful discussions. This work was supported by JSPS KAKENHI Grant Nos. 18K03525 and 21H05185 (D.Y.) and JST PRESTO Grant No. JPMJPR2118, Japan (D.Y.).
\end{acknowledgments}

\end{document}


\maketitle

\section{PLAQUETTE MEAN-FIELD APPROXIMATION}
Let us rewrite the Hamiltonian $\hat{\mathcal{H}}=\hat{\mathcal{H}}_{\rm SU(4)}+\hat{\mathcal{H}}_D$ in the main text as follows:
\begin{eqnarray}
\hat{\mathcal{H}}&=&J\sum_{I}^{N/4}\sum_{\braket{n,n'}}\sum_{\mu,\nu}\hat{F}^{\nu}_{\mu}(I_n)\hat{F}^{\mu}_{\nu}(I_{n'})+\sum_{I}^{N/4}\sum_{I'}^{N/4}\sum_{n,n'=1}^4\sum_{\mu,\nu}\mathcal{J}'_{I_n,I'_{n'}}
\hat{F}^{\nu}_{\mu}(I_n)\hat{F}^{\mu}_{\nu}(I'_{n'})\notag\\
& & +D\sum_I^{N/4}\sum_{n=1}^4\left[\hat{F}^{\rm c}_{\rm c}(I_n)+\hat{F}^{\rm s}_{\rm s}(I_n)-\hat{F}^{\rm u}_{\rm u}(I_n)-\hat{F}^{\rm d}_{\rm d}(I_n)\right].
\end{eqnarray}
$N$ is the total number of sites and $I_n$ denotes a site  in  plaquette $I$: $I_1$ for the top-left, $I_2$ for the top-right, $I_3$ for the bottom-right, and $I_4$ for the  bottom-left (clock-wise convention). $\braket{n,n'}$ denotes the pairs of neighboring sites in a plaquette.
We define
\begin{eqnarray}
\mathcal{J}'_{I_n,I'_{n'}}=
\left\{
    \begin{array}{cl}
      J' & \ \ \ {\rm if}\ |\mbox{\boldmath$r$}_{I'_{n'}}-\mbox{\boldmath$r$}_{I_{n}}|=1\ {\rm and}\ I\neq I'\\
      0 & \ \ \ \rm{otherwise}\\
    \end{array}
\right. .
\end{eqnarray}
In order to reduce the full problem 
on  the tetramerized square lattice shown in Fig. 1 to a four-site one, we apply the mean-field approximation $\hat{F}^{\nu}_{\mu}(I_n)\hat{F}^{\mu}_{\nu}(I'_{n'})\approx
\mathcal{h}\hat{F}^{\nu}_{\mu}(I_n)\mathcal{i}\hat{F}^{\mu}_{\nu}(I'_{n'})+
\mathcal{h}\hat{F}^{\mu}_{\nu}(I'_{n'})\mathcal{i}\hat{F}^{\nu}_{\mu}(I_n)-
\mathcal{h}\hat{F}^{\nu}_{\mu}(I_n)\mathcal{i}\mathcal{h}\hat{F}^{\mu}_{\nu}(I'_{n'})\mathcal{i}$ to the interplaquette bonds.  
We obtain the following plaquette mean-field (PMF) Hamiltonian:
\begin{eqnarray}
\hat{\mathcal{H}}^{\textup{PMF}}_I&=&
\sum_{\braket{n,n'}}\sum_{\mu,\nu}\hat{F}^{\nu}_{\mu}(I_n)\hat{F}^{\mu}_{\nu}(I_{n'})\notag\\
& &+\sum_{I'}^{N/4}\sum_{n,n'=1}^4\sum_{\mu,\nu}\mathcal{J}'_{I_n,I'_{n'}}
\left[\hat{F}^{\nu}_{\mu}(I_n)\braket{\hat{F}^{\mu}_{\nu}(I'_{n'})}+\hat{F}^{\mu}_{\nu}(I'_{n'})\braket{\hat{F}^{\nu}_{\mu}(I_n)}-\braket{\hat{F}^{\nu}_{\mu}(I_n)}\braket{\hat{F}^{\mu}_{\nu}(I'_{n'})}\right]\notag\\
& & +D\sum_{n=1}^4\left[\hat{F}^{\rm c}_{\rm c}(I_n)+\hat{F}^{\rm s}_{\rm s}(I_n)-\hat{F}^{\rm u}_{\rm u}(I_n)-\hat{F}^{\rm d}_{\rm d}(I_n)\right].
\end{eqnarray}
 Under the assumption of the plaquette-checkerboard pattern of the mean-field parameters shown in Fig. 1, we rewrite $\hat{\mathcal{H}}$ as $\hat{\mathcal{H}}=\sum_{I\in{\rm A}}\hat{\mathcal{H}}^{\rm{PMF}}_{I}+\sum_{I\in{\rm B}}\hat{\mathcal{H}}^{\rm{PMF}}_{I}$.
At this stage we are thus neglecting the interplaquette quantum entanglement.
 We self-consistently calculate the mean-field parameters  $\{\braket{\hat{F}^\nu_\mu(I_n\in{\rm A})}\}$ and $\{\braket{\hat{F}^\nu_\mu(I_n\in{\rm B})}\}$ for $i=1,2,3,4$ and $\nu,\mu\in\{{\rm c,u,d,s}\}$. The values of the mean-field parameters of SU(4) singlet and RVB${}_{\rm ud}$ phase, in which $\{\braket{\hat{F}^\nu_\mu(I_n\in{\rm A})}\}=\{\braket{\hat{F}^\nu_\mu(I_n\in{\rm B})}\}$ is satisfied,  are shown in Tables~\ref{tab:aeqb1} and \ref{tab:aeqb2}

\begin{table}[h]
    \caption{Mean-field parameters $\braket{\hat{F}^{\nu}_{\mu}(i)}$ of SU(4) singlet phase}
        \centering
        \begin{tabular}{|c|c|c|c|c|c|c|c|c|c|c|c|c|c|c|c|c|c|c|c|c|}\hline
        $n$ &  1 & 2 & 3 & 4    & 1 & 2 & 3 & 4 & 1 & 2 & 3 & 4 & 1 & 2 & 3 & 4\\
        \hline
       \diagbox{$\nu$}{$\mu$} &\multicolumn{4}{|c|}{c} & \multicolumn{4}{|c|}{u} & \multicolumn{4}{|c|}{d} &\multicolumn{4}{|c|}{s} \\
       \hline
     c &  $\frac{1}{4}$ & $\frac{1}{4}$ & $\frac{1}{4}$ & $\frac{1}{4}$   & $0$ & $0$ & $0$ & $0$ & $0$ & $0$ & $0$ & $0$ & $0$ & $0$ & $0$ & $0$\\
       \hline
     u &  $0$ & $0$ & $0$ & $0$ & $\frac{1}{4}$ & $\frac{1}{4}$ & $\frac{1}{4}$ & $\frac{1}{4}$   &  $0$ & $0$ & $0$ & $0$ & $0$ & $0$ & $0$ & $0$\\
       \hline
     d &  $0$ & $0$ & $0$ & $0$ & $0$ & $0$ & $0$ & $0$ & $\frac{1}{4}$ & $\frac{1}{4}$ & $\frac{1}{4}$ & $\frac{1}{4}$   &   $0$ & $0$ & $0$ & $0$\\
       \hline
     s &  $0$ & $0$ & $0$ & $0$ & $0$ & $0$ & $0$ & $0$ & $0$ & $0$ & $0$ & $0$ &$\frac{1}{4}$ & $\frac{1}{4}$ & $\frac{1}{4}$ & $\frac{1}{4}$\\
       \hline
\end{tabular}
\label{tab:aeqb1}
\end{table}

\begin{table}[h]
    \caption{Mean-field parameters $\braket{\hat{F}^{\nu}_{\mu}(i)}$ of RVB${}_{\rm ud}$ phase}
        \centering
        \begin{tabular}{|c|c|c|c|c|c|c|c|c|c|c|c|c|c|c|c|c|c|c|c|c|}\hline
        $n$ &  1 & 2 & 3 & 4    & 1 & 2 & 3 & 4 & 1 & 2 & 3 & 4 & 1 & 2 & 3 & 4\\
        \hline
       \diagbox{$\nu$}{$\mu$} &\multicolumn{4}{|c|}{c} & \multicolumn{4}{|c|}{u} & \multicolumn{4}{|c|}{d} &\multicolumn{4}{|c|}{s} \\
       \hline
       c &  $0$ & $0$ & $0$ & $0$   & $0$ & $0$ & $0$ & $0$ & $0$ & $0$ & $0$ & $0$ & $0$ & $0$ & $0$ & $0$\\
       \hline
       u &  $0$ & $0$ & $0$ & $0$ & $\frac{1}{2}$ & $\frac{1}{2}$ & $\frac{1}{2}$ & $\frac{1}{2}$   &  $0$ & $0$ & $0$ & $0$ & $0$ & $0$ & $0$ & $0$\\
       \hline
       d &  $0$ & $0$ & $0$ & $0$ & $0$ & $0$ & $0$ & $0$ & $\frac{1}{2}$ & $\frac{1}{2}$ & $\frac{1}{2}$ & $\frac{1}{2}$   &   $0$ & $0$ & $0$ & $0$\\
       \hline
       s &  $0$ & $0$ & $0$ & $0$ & $0$ & $0$ & $0$ & $0$ & $0$ & $0$ & $0$ & $0$ &$0$ & $0$ & $0$ & $0$\\
       \hline
\end{tabular}
\label{tab:aeqb2}
\end{table}

The singlet-RVB${}_{\rm ud}$ solid phase is characterized by  by $\{\braket{\hat{F}^\nu_\mu(I_n\in{\rm A})}\}$ as in Table~\ref{tab:aeqb1} and $\{\braket{\hat{F}^\nu_\mu(I_n\in{\rm B})}\}$ as in Table~\ref{tab:aeqb2}.

\clearpage
\section{LINEAR FLAVOR-WAVE THEORY}
For convenience, we write the Hamiltonian as the sum of three terms, namely $\hat{\mathcal{H}}=\hat{\mathcal{H}}^{(J)}+\hat{\mathcal{H}}^{(J')}+\hat{\mathcal{H}}^{(D)}$, where
\begin{eqnarray}
\hat{\mathcal{H}}^{(J)}&=&J\sum_{I}^{N/4}\sum_{\braket{n,n'}}\sum_{\mu,\nu}\hat{F}^{\nu}_{\mu}(I_n)\hat{F}^{\mu}_{\nu}(I_{n'}),\\
\hat{\mathcal{H}}^{(J')}&=&\sum_{I}^{N/4}\sum_{I'}^{N/4}\sum_{n,n'=1}^4\sum_{\mu,\nu}\mathcal{J}'_{I_n,I'_{n'}}
\hat{F}^{\nu}_{\mu}(I_n)\hat{F}^{\mu}_{\nu}(I'_{n'}),\\
\hat{\mathcal{H}}^{(D)}&=&D\sum_I^{N/4}\sum_{n=1}^4\left[\hat{F}^{\rm c}_{\rm c}(I_n)+\hat{F}^{\rm s}_{\rm s}(I_n)-\hat{F}^{\rm u}_{\rm u}(I_n)-\hat{F}^{\rm d}_{\rm d}(I_n)\right].
\end{eqnarray}

For simplicity, here we describe in detail the next steps of the procedure for the intraplaquette hopping term $\hat{\mathcal{H}}^{(J)}$, but the same applies to the other two terms. Let $\mbox{\boldmath$d$}_{I}$ and $\tilde{\mbox{\boldmath$d$}}_{I}$ be the  column vectors whose components are the annihilation  operators of the  256 original basis states of a plaquette  $\{\hat{a}_{I,l}\}$ and of the PMF eigenstates,
respectively:
\begin{equation}
		\mbox{\boldmath$d$}_{I}\equiv
        \left(
		\begin{array}{cc}
		\hat{a}_{I,1}\\
        \hat{a}_{I,2}\\
        \vdots\\
		\hat{a}_{I,256}
		\end{array}\right),\ \ \ 
        \tilde{\mbox{\boldmath$d$}}_{I}\equiv
        \left(
		\begin{array}{cc}
		\hat{g}_{I}\\
        \hat{e}_{I,1}\\
        \vdots\\
		\hat{e}_{I,255}
		\end{array}\right)= \left(
		\begin{array}{cc}
		\hat{g}_{I}\\
         \hline\\
        \mbox{\boldmath$e$}_{I}\\
		\\
		\end{array}\right),
\end{equation}
where $\hat{g}_I$ and $\{\hat{e}_{I,m}\}$ denote the annihilation operators of the ground state and the excited states at the PMF level, respectively, and $I$ is plaquette index. $\mbox{\boldmath$d$}_{I}$ and $\tilde{\mbox{\boldmath$d$}}_{I}$ are related by a $256\times256$ unitary matrix as  $\tilde{\mbox{\boldmath$d$}}_{I}=U_{I}\mbox{\boldmath$d$}_{I}$.
We thus can represent the Hamiltonian 
as
\begin{eqnarray}
\hat{\mathcal{H}}^{(J)}&=&J\sum_{I}^{N/4}\sum_{\braket{n,n'}}\sum_{\mu,\nu}\mbox{\boldmath$d$}_{I}^\dagger \hat{F}^{\nu}_{\mu}(I_n)\hat{F}^{\mu}_{\nu}(I_{n'})\mbox{\boldmath$d$}_{I}\notag\\
&=&J\sum_{I}^{N/4}\sum_{\braket{n,n'}}\sum_{\mu,\nu}\tilde{\mbox{\boldmath$d$}}_{I}^\dagger U_{I}\hat{F}^{\nu}_{\mu}(I_n)\hat{F}^{\mu}_{\nu}(I_{n'})U_{I}^\dagger\tilde{\mbox{\boldmath$d$}}_{I}.
\end{eqnarray}
The matrix $\sum_{\braket{n,n'}}\sum_{\mu,\nu}U_{I}\hat{F}^{\nu}_{\mu}(I_n)\hat{F}^{\mu}_{\nu}(I_{n'})U_{I}^\dagger$ has block form:
\clearpage
\begin{figure}[h]
    \centering
    \includegraphics[width=12cm]{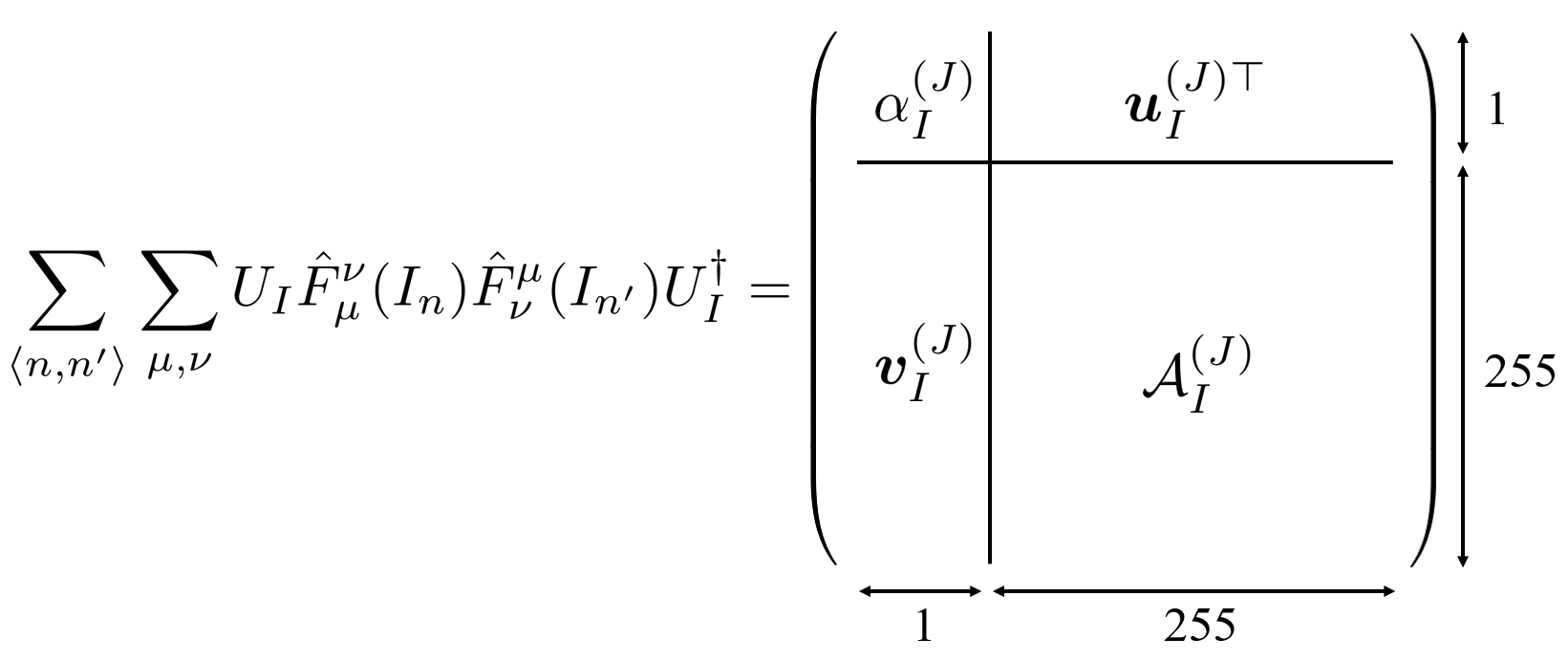}
\label{f6}
\end{figure}

From this we obtain
\begin{eqnarray}
\hat{\mathcal{H}}^{(J)}=J\left(\sum_{I\in\rm{A}}^{N/8}+\sum_{I\in\rm{B}}^{N/8}\right)\left[
\alpha_{I}^{(J)}\hat{g}_{I}^\dagger\hat{g}_{I}+\hat{g}_{I}^\dagger\mbox{\boldmath$u$}_{I}^{(J)\top}\cdot\mbox{\boldmath$e$}_{I}+\mbox{\boldmath$e$}_{I}^\dagger\cdot\mbox{\boldmath$v$}_{I}^{(J)}\hat{g}_{I}+\mbox{\boldmath$e$}_{I}^\dagger\mathcal{A}_{I}^{(J)}\mbox{\boldmath$e$}_{I}
\right].
\end{eqnarray}
$\hat{g}_I$ and $\{\hat{e}_{I,m}\}$ obey the bosonic commutation relations under the constraint    
\begin{equation}
 \hat{g}_I^{\dagger} \hat{g}_I+\sum_{m}\hat{e}_{I,m}^{\dagger} \hat{e}_{I,m}=\hat{g}_I^\dagger\hat{g}_I+\mbox{\boldmath$e$}_I^\dagger\cdot\mbox{\boldmath$e$}_I=1.
 \end{equation}
Following the analogy with the spin-wave expansion method, we may expand the Hamiltonian with 
 \begin{equation}
 \hat{g}_I^\dagger=\sqrt{1-\mbox{\boldmath$e$}_I^\dagger\cdot\mbox{\boldmath$e$}}\simeq1-\frac{1}{2}\mbox{\boldmath$e$}_I^\dagger\cdot\mbox{\boldmath$e$}+\cdots,
 \end{equation}
under the assumption that the fluctuations $\mbox{\boldmath$e$}_I^\dagger\cdot\mbox{\boldmath$e$}_I$ around the PMF ground state are sufficiently small.
The Hamiltonian becomes
\begin{equation}
    \hat{\mathcal{H}}^{(J)}=E_0^{(J)}+\hat{\mathcal{H}}_1^{(J)}+\hat{\mathcal{H}}_2^{(J)}+\hat{\mathcal{H}}_3^{(J)}+\cdots,
\end{equation}
where 
\begin{eqnarray}
E_0^{(J)}&=&\alpha_I^{(J)}\\
\hat{\mathcal{H}}^{(J)}_1&=&\mbox{\boldmath$u$}_{I}^{(J)\top}\cdot\mbox{\boldmath$e$}_{I}+\mbox{\boldmath$e$}_{I}^\dagger\cdot\mbox{\boldmath$v$}_{I}^{(J)}\\
\hat{\mathcal{H}}^{(J)}_2&=&-\alpha_I^{(J)}\mbox{\boldmath$e$}_{I}^\dagger\cdot\mbox{\boldmath$e$}_{I}+\mbox{\boldmath$e$}_{I}^\dagger\mathcal{A}_{I}^{(J)}\mbox{\boldmath$e$}_{I}\\
\hat{\mathcal{H}}^{(J)}_3&=&-\frac{1}{2}\left(\mbox{\boldmath$e$}_{I}^\dagger\cdot\mbox{\boldmath$e$}_{I}\mbox{\boldmath$u$}_{I}^{(J)\top}\cdot\mbox{\boldmath$e$}_{I}+\mbox{\boldmath$e$}_{I}^\dagger\cdot\mbox{\boldmath$v$}_{I}^{(J)}\mbox{\boldmath$e$}_{I}^\dagger\cdot\mbox{\boldmath$e$}_{I}\right)
\end{eqnarray}
Note that $\hat{\mathcal{H}}_1^{(J)}=0$ is justified when we consider the full Hamiltonian. Following the assumption of the plaquette-checkerboard structure (A-plaquette and B-plaquette), we define the following Fourier transformation
\begin{eqnarray}
    \mbox{\boldmath$e$}_{I\in\rm{A}}&=&\frac{1}{\sqrt{N/8}}\sum_{\mbox{\boldmath$k$}\in\rm{B.Z.}}\mbox{\boldmath$e$}^{(\rm{A})}_{\mbox{\boldmath$k$}}\exp(i\mbox{\boldmath$k$}\cdot\mbox{\boldmath$r$}_{I\in\rm{A}})\\
    \mbox{\boldmath$e$}_{I\in\rm{B}}&=&\frac{1}{\sqrt{N/8}}\sum_{\mbox{\boldmath$k$}\in\rm{B.Z.}}\mbox{\boldmath$e$}^{(\rm{B})}_{\mbox{\boldmath$k$}}\exp(i\mbox{\boldmath$k$}\cdot\mbox{\boldmath$r$}_{I\in\rm{B}})
\end{eqnarray}
where $\mbox{\boldmath$r$}_{I}$ is the position vector of the center of the plaquette $I$. We obtain
\begin{eqnarray}
E^{(J)}_0+\hat{\mathcal{H}}^{(J)}_2&=&
J\sum_{\mbox{\boldmath$k$}}
\left[-\alpha_{I\in{\rm{A}}}^{(J)}\mbox{\boldmath$e$}^{(\rm{A})\dagger}_{\mbox{\boldmath$k$}}\cdot\mbox{\boldmath$e$}^{(\rm{A})}_{\mbox{\boldmath$k$}}+\mbox{\boldmath$e$}^{(\rm{A})\dagger}_{\mbox{\boldmath$k$}}\mathcal{A}_{I\in{\rm{A}}}^{(J)}\mbox{\boldmath$e$}^{(\rm{A})}_{\mbox{\boldmath$k$}}\right.\notag\\
& &\left.-\alpha_{I\in{\rm{B}}}^{(J)}\mbox{\boldmath$e$}^{(\rm{B})\dagger}_{\mbox{\boldmath$k$}}\cdot\mbox{\boldmath$e$}^{(\rm{B})}_{\mbox{\boldmath$k$}}+\mbox{\boldmath$e$}^{(\rm{B})\dagger}_{\mbox{\boldmath$k$}}\mathcal{A}_{I\in{\rm{B}}}^{(J)}\mbox{\boldmath$e$}^{(\rm{B})}_{\mbox{\boldmath$k$}}\right]+\frac{1}{8}JN(\alpha_{I\in{\rm{A}}}^{(J)}+\alpha_{I\in{\rm{B}}}^{(J)}).
\end{eqnarray}
Applying the same procedure to the other terms we arrive at
\begin{eqnarray}
E^{(D)}_0+\hat{\mathcal{H}}^{(D)}_2&=&
J\sum_{\mbox{\boldmath$k$}}
\left[-\alpha_{I\in{\rm{A}}}^{(D)}\mbox{\boldmath$e$}^{(\rm{A})\dagger}_{\mbox{\boldmath$k$}}\cdot\mbox{\boldmath$e$}^{(\rm{A})}_{\mbox{\boldmath$k$}}+\mbox{\boldmath$e$}^{(\rm{A})\dagger}_{\mbox{\boldmath$k$}}\mathcal{A}_{I\in{\rm{A}}}^{(D)}\mbox{\boldmath$e$}^{(\rm{A})}_{\mbox{\boldmath$k$}}\right.\notag\\
& &\left.-\alpha_{I\in{\rm{B}}}^{(D)}\mbox{\boldmath$e$}^{(\rm{B})\dagger}_{\mbox{\boldmath$k$}}\cdot\mbox{\boldmath$e$}^{(\rm{B})}_{\mbox{\boldmath$k$}}+\mbox{\boldmath$e$}^{(\rm{B})\dagger}_{\mbox{\boldmath$k$}}\mathcal{A}_{I\in{\rm{B}}}^{(D)}\mbox{\boldmath$e$}^{(\rm{B})}_{\mbox{\boldmath$k$}}\right]+\frac{1}{8}DN(\alpha_{I\in{\rm{A}}}^{(D)}+\alpha_{I\in{\rm{B}}}^{(D)}),
\end{eqnarray}
\begin{eqnarray}
E^{(J')}_0+\hat{\mathcal{H}}^{(J')}_2&=&\sum_{I\in\rm{A}}^{N/8}\sum_{n,n'=1}^{4}\sum_{\mu,\nu}\sum_{\mbox{\boldmath$k$}}\frac{1}{2}\mathcal{J}'_{I_n,I'_{n'}}\left[
-\alpha_{I_n}^{(J')}{}^\nu_\mu\alpha_{I'_{n'}}^{(J')}{}^\mu_\nu
\mbox{\boldmath$e$}^{(\rm{B})\dagger}_{\mbox{\boldmath$k$}}\cdot\mbox{\boldmath$e$}^{(\rm{B})}_{\mbox{\boldmath$k$}}+\alpha_{I_n}^{(J')}{}^\nu_\mu\mbox{\boldmath$e$}^{(\rm{B})\dagger}_{\mbox{\boldmath$k$}}\mathcal{A}_{I'_{n'}}^{(J')}{}^\mu_\nu\mbox{\boldmath$e$}^{(\rm{B})}_{\mbox{\boldmath$k$}}\right]+\notag\\
& &
\sum_{I'\in\rm{B}}^{N/8}\sum_{n,n'=1}^{4}\sum_{\mu,\nu}\sum_{\mbox{\boldmath$k$}}\left[\frac{1}{2}\mathcal{J}'_{I_n,I'_{n'}}\left(-\alpha_{I_n}^{(J')}{}^\nu_\mu\alpha_{I'_{n'}}^{(J')}{}^\mu_\nu
\mbox{\boldmath$e$}^{(\rm{A})\dagger}_{\mbox{\boldmath$k$}}\cdot\mbox{\boldmath$e$}^{(\rm{A})}_{\mbox{\boldmath$k$}}+\alpha_{I'_{n'}}^{(J')}{}^\mu_\nu\mbox{\boldmath$e$}^{(\rm{A})\dagger}_{\mbox{\boldmath$k$}}\mathcal{A}_{I_{n}}^{(J')}{}^\nu_\mu\mbox{\boldmath$e$}^{(\rm{A})}_{\mbox{\boldmath$k$}}\right)\right.\notag\\
& &
+2\gamma_{-\mbox{\boldmath$k$},I_n,I'_{n'}}\left\{\mbox{\boldmath$e$}^{(\rm{A})\top}_{\mbox{\boldmath$k$}}\left(\mbox{\boldmath$u$}_{I_{n}}^{(J')}{}^\nu_\mu\otimes\mbox{\boldmath$u$}_{I'_{n'}}^{(J')}{}^\mu_\nu{}^\top\right)\mbox{\boldmath$e$}^{(\rm{B})}_{-\mbox{\boldmath$k$}}
+\mbox{\boldmath$e$}^{(\rm{A})\top}_{\mbox{\boldmath$k$}}\left(\mbox{\boldmath$u$}_{I_{n}}^{(J')}{}^\nu_\mu\otimes\mbox{\boldmath$v$}_{I'_{n'}}^{(J')}{}^\mu_\nu{}^\top\right)\mbox{\boldmath$e$}^{(\rm{B})*}_{\mbox{\boldmath$k$}}\right\}\notag\\
& &
\left.+2\gamma_{\mbox{\boldmath$k$},I_n,I'_{n'}}\left\{\mbox{\boldmath$e$}^{(\rm{A})\dagger}_{\mbox{\boldmath$k$}}\left(\mbox{\boldmath$v$}_{I_{n}}^{(J')}{}^\nu_\mu\otimes\mbox{\boldmath$u$}_{I'_{n'}}^{(J')}{}^\mu_\nu{}^\top\right)\mbox{\boldmath$e$}^{(\rm{B})}_{\mbox{\boldmath$k$}}
+\mbox{\boldmath$e$}^{(\rm{A})\dagger}_{\mbox{\boldmath$k$}}\left(\mbox{\boldmath$v$}_{I_{n}}^{(J')}{}^\nu_\mu\otimes\mbox{\boldmath$v$}_{I'_{n'}}^{(J')}{}^\mu_\nu{}^\top\right)\mbox{\boldmath$e$}^{(\rm{B})*}_{-\mbox{\boldmath$k$}}\right\}\right]\notag\\
& &+
\sum_{I\in\rm{A}}^{N/8}\sum_{I'\in\rm{B}}^{N/8}\sum_{n,n'=1}^{4}\sum_{\mu,\nu}\frac{1}{2}\mathcal{J}'_{I_n,I'_{n'}}\alpha_{I_n}^{(J')}{}^\nu_\mu\alpha_{I'_{n'}}^{(J')}{}^\mu_\nu+(\rm{A}\leftrightarrow\rm{B}),
\end{eqnarray}
where
\begin{equation}
\gamma_{\mbox{\boldmath$k$},I_n,I'_{n'}}=\frac{1}{4}\mathcal{J}'_{I_n,I'_{n'}}\exp\left[i(\mbox{\boldmath$r$}_{I'}-\mbox{\boldmath$r$}_{I})\cdot\mbox{\boldmath$k$}\right].
\end{equation}
Passing to the Bogoliubov matrix representation of the full Hamiltonian  we get
\begin{equation}
   \hat{\mathcal{H}}=\frac{1}{2}\sum_{\mbox{\boldmath$k$}}
    \left(
    \begin{array}{cccc}
     \mbox{\boldmath$e$}^{(\rm{A})\dagger}_{\mbox{\boldmath$k$}} & \mbox{\boldmath$e$}^{(\rm{B})\dagger}_{\mbox{\boldmath$k$}} & \mbox{\boldmath$e$}^{(\rm{A})\top}_{-\mbox{\boldmath$k$}} & \mbox{\boldmath$e$}^{(\rm{B})\top}_{-\mbox{\boldmath$k$}}
    \end{array}
  \right)
    H_{\mbox{\boldmath$k$}} 
  \left(
    \begin{array}{cccc}
      \mbox{\boldmath$e$}^{(\rm{A})}_{\mbox{\boldmath$k$}}  \\
      \mbox{\boldmath$e$}^{(\rm{B})}_{\mbox{\boldmath$k$}}  \\
      \mbox{\boldmath$e$}^{(\rm{A})*}_{-\mbox{\boldmath$k$}}  \\
      \mbox{\boldmath$e$}^{(\rm{B})*}_{-\mbox{\boldmath$k$}}
    \end{array}
  \right)+E_0-\frac{1}{2}\sum_{\mbox{\boldmath$k$}}\textup{Tr}A_{-\mbox{\boldmath$k$}}^*,
    \end{equation}
where $H_{\mbox{\boldmath$k$}}$ is the $1020\times1020$ Hermitian matrix made up of  four square blocks: 
    \begin{eqnarray}
H_{\mbox{\boldmath$k$}}=\left(
    \begin{array}{cc}
      A_{\mbox{\boldmath$k$}} & B_{\mbox{\boldmath$k$}}\\
      B_{-\mbox{\boldmath$k$}}^* & A_{-\mbox{\boldmath$k$}}^*
    \end{array}
  \right),
\end{eqnarray}
\begin{eqnarray}
A_{\mbox{\boldmath$k$}}&=&\left(
    \begin{array}{cc}
      (-J\alpha_{I_{\rm{A}}}^{(J)}-D\alpha_{I_{\rm{A}}}^{(D)})\mathbbm{1}+J\mathcal{A}_{I_{\rm{A}}}^{(J)}+D\mathcal{A}_{I_{\rm{A}}}^{(D)} & \mathbb{O}\\
      \mathbb{O} & (-J\alpha_{I_{\rm{B}}}^{(J)}-D\alpha_{I_{\rm{A}}}^{(D)})\mathbbm{1}+J\mathcal{A}_{I_{\rm{B}}}^{(J)}+D\mathcal{A}_{I_{\rm{B}}}^{(D)}
    \end{array}
  \right)+\notag\\
  & &\sum_{\mu,\nu}\sum_{n,n'=1}^4\left[
  \sum_{I'\in\rm{B}}^{N/8}\left(
    \begin{array}{cc}
      \Gamma_1+\Gamma_2 & \Gamma_4\\
      \mathbb{O} & \mathbb{O}
    \end{array}
  \right)+
  \sum_{I\in\rm{A}}^{N/8}\left(
    \begin{array}{cc}
      \mathbb{O} & \mathbb{O}\\
      \mathbb{O} & \Gamma_1+\Gamma_3
    \end{array}
  \right)+
  \sum_{I'\in\rm{A}}^{N/8}\left(
    \begin{array}{cc}
      \mathbb{O} & \mathbb{O}\\
      \Gamma_4 & \mathbb{O}
    \end{array}
  \right)\right]\\
  B_{\mbox{\boldmath$k$}}&=&
  \sum_{\mu,\nu}\sum_{n,n'=1}^4\left[
  \sum_{I'\in\rm{B}}^{N/8}\left(
    \begin{array}{cc}
      \mathbb{O} & \Gamma_5\\
      \mathbb{O} & \mathbb{O}
    \end{array}
  \right)+
  \sum_{I'\in\rm{A}}^{N/8}\left(
    \begin{array}{cc}
      \mathbb{O} & \mathbb{O}\\
      \Gamma_5 & \mathbb{O}
    \end{array}
  \right)\right]
\end{eqnarray}
\begin{equation}
    \left\{
    \begin{array}{l}
      \Gamma_1=-\mathcal{J}'_{I_n,I'_{n'}}\alpha_{I_n}^{(J')}{}^\nu_\mu\alpha_{I'_{n'}}^{(J')}{}^\mu_\nu \mathbbm{1}\\
      \Gamma_2=\mathcal{J}'_{I_n,I'_{n'}}\alpha_{I'_{n'}}^{(J')}{}^\mu_\nu\mathcal{A}_{I_{n}}^{(J')}{}^\nu_\mu\\
\Gamma_3=\mathcal{J}'_{I_n,I'_{n'}}\alpha_{I_{n}}^{(J')}{}^\mu_\nu\mathcal{A}_{I'_{n'}}^{(J')}{}^\nu_\mu\\
\Gamma_4=4\gamma_{\mbox{\boldmath$k$},I_n,I'_{n'}}\left(\mbox{\boldmath$v$}_{I_{n}}^{(J')}{}^\nu_\mu\otimes\mbox{\boldmath$u$}_{I'_{n'}}^{(J')}{}^\mu_\nu{}^\top\right)\\
\Gamma_5=4\gamma_{\mbox{\boldmath$k$},I_n,I'_{n'}}\left(\mbox{\boldmath$v$}_{I_{n}}^{(J')}{}^\nu_\mu\otimes\mbox{\boldmath$v$}_{I'_{n'}}^{(J')}{}^\mu_\nu{}^\top\right)
    \end{array}
  \right..
\end{equation}
Here, $\mathbbm{1}$ and $\mathbb{O}$ are the $255\times255$ identity matrix and zero matrix, respectively. $H_{\mbox{\boldmath$k$}}$ is diagonalized by a Bogoliubov transformation
 \begin{equation}
 \left(
    \begin{array}{c}
      \mbox{\boldmath$e$}_{\mbox{\boldmath$k$}}^{(\rm{A})}\\
      \mbox{\boldmath$e$}_{\mbox{\boldmath$k$}}^{(\rm{B})}\\
      \mbox{\boldmath$e$}_{-\mbox{\boldmath$k$}}^{(\rm{A})*}\\
      \mbox{\boldmath$e$}_{-\mbox{\boldmath$k$}}^{(\rm{B})*}\\
    \end{array}
  \right)=U^{\rm{P}}_{\mbox{\boldmath$k$}}
  \left(
    \begin{array}{c}
      \hat{\varepsilon}_{\mbox{\boldmath$k$},1}\\
      \vdots\\
      \hat{\varepsilon}_{\mbox{\boldmath$k$},510}\\
      \hat{\varepsilon}_{-\mbox{\boldmath$k$},1}^\dagger\\
      \vdots\\
      \hat{\varepsilon}_{-\mbox{\boldmath$k$},510}^\dagger\\
    \end{array}
  \right)
 \end{equation}
 where $U^{\rm{P}}_{\mbox{\boldmath$k$}}$ is the $1020\times1020$  paraunitary matrix. We obtain the diagonalized form as
 \begin{equation}
    \hat{\mathcal{H}}_2=\sum_{\mbox{\boldmath$k$}}\sum_{\lambda=1}^{510}\omega_{\mbox{\boldmath$k$},\lambda}\hat{\varepsilon}^\dagger_{\mbox{\boldmath$k$},\lambda}\hat{\varepsilon}_{\mbox{\boldmath$k$},\lambda}+E_{\rm{zp}},
\end{equation}
where $\omega_{\mbox{\boldmath$k$},\lambda}$ is the $\lambda$-th excitation energy band and 
\begin{equation}
E_{\rm{zp}}=\frac{1}{2}\sum_{\mbox{\boldmath$k$}}\sum_{\lambda}\omega_{\mbox{\boldmath$k$},\lambda}-\frac{1}{2}\sum_{\mbox{\boldmath$k$}}\textup{Tr}A_{-\mbox{\boldmath$k$}}^*.
\end{equation}

\newpage
\section{FURTHER COMMENTS ON THE VALIDITY OF THE LFWT APPROACH}
\begin{figure}[h]
    \centering
    \includegraphics[width=13cm]{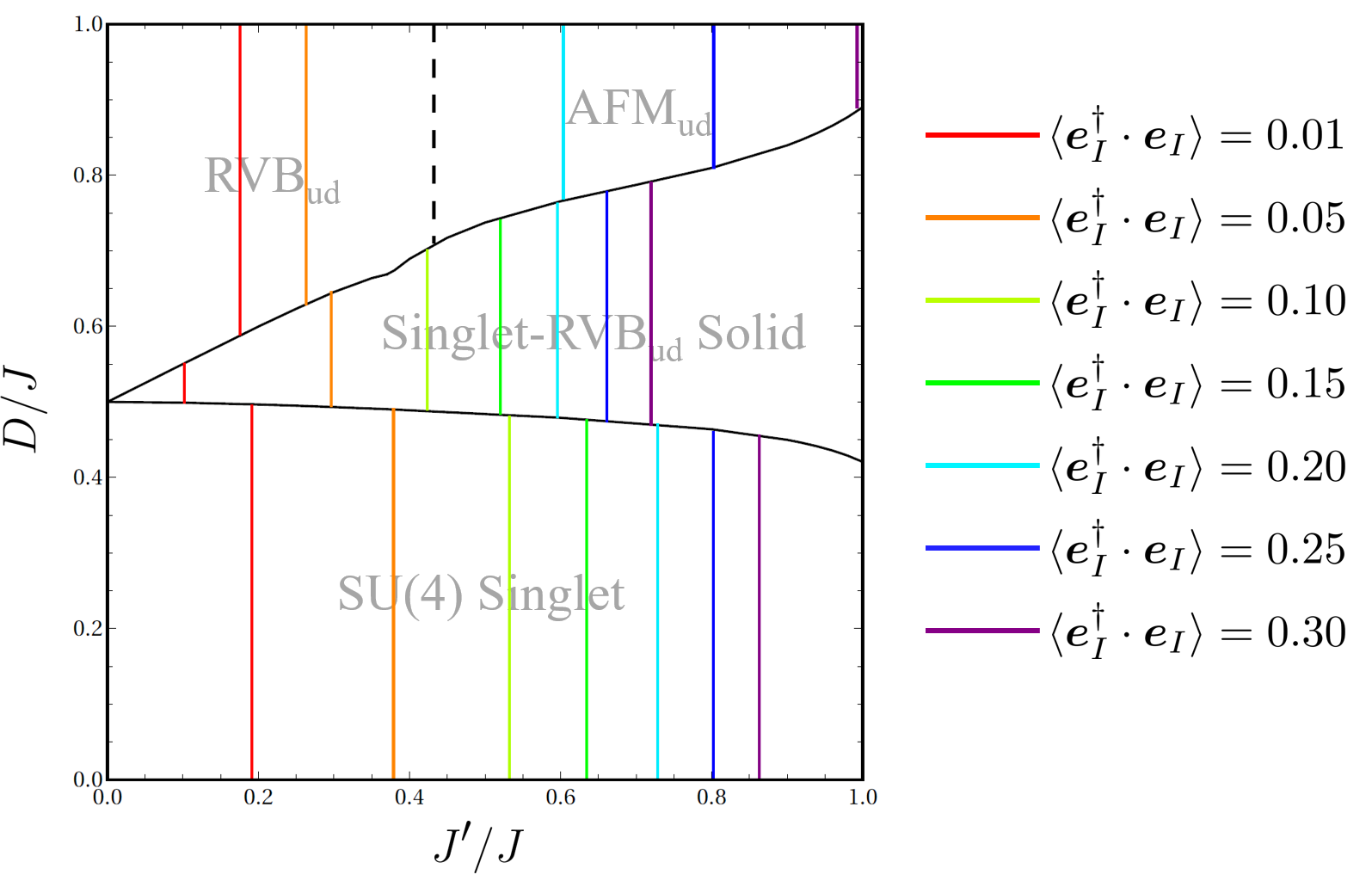}
    \caption{Constant-$\braket{\mbox{\boldmath$e$}_I^\dagger\cdot\mbox{\boldmath$e$}_I}$ lines within the various phases.}
   \label{ee}
\end{figure}

In the main text we pointed out that our LFWT approach is more reliable in the regime of small $J'/J$ because it is based on the PMF approximation. Here we discuss another aspect, namely the validity of the expansion of the square root in Eq.~4 in the main text and the subsequent truncation that leads to the Bogoliubov Hamiltonian.  Such procedure can be justified only when the occupation of the excited states $\braket{\mbox{\boldmath$e$}_I^\dagger\cdot\mbox{\boldmath$e$}_I}=\sum_{n}\braket{\hat{e}_{I,n}^\dagger\hat{e}_{I,n}}$ is small. In Fig.~\ref{ee} we show various constant-$\braket{\mbox{\boldmath$e$}_I^\dagger\cdot\mbox{\boldmath$e$}_I}$ lines within the different phases in the phase diagram. For the states with a two-sublattice structure, we picked the larger one from $\braket{\mbox{\boldmath$e$}_I^\dagger\cdot\mbox{\boldmath$e$}_I}$ for $I\in$ A and $I\in$ B. As can be seen this quantity is independent of $D$ (within our approximations) except for jumps at phase boundaries and, most importantly, it is sufficiently small in a large region of the phase diagram; in particular it does not exceed 0.1 for $J'/J < 0.4$. This analysis provides an {\it a posteriori} justification of the LFWT treatment, especially in the region of the parameter space where also the assumptions of the PMF approximation are more under control. The AFM phase has a relatively higher value of $\braket{\mbox{\boldmath$e$}_I^\dagger\cdot\mbox{\boldmath$e$}_I}$, which is of course expected and related to the well-known magnetic moment depletion of quantum antiferromagnets.

